# Present After Presence:

## Subtraction, Givenness, and the Structure of Being-There

**TOIDA Koichi**
MESON, Inc.
koichi.toida@meson.tokyo

**ABSTRACT** Presence research has often proceeded by addition, treating immersion, embodiment, agency, ownership, co-presence, reciprocity, and temporal simultaneity as conditions that stabilise the sense of being there. Yet immersive video suggests that several of these conditions can be weakened without eliminating Presence. Building on Bodyless Presence and Bodyless Presentness, this paper asks what is disclosed when such conditions are progressively subtracted. I distinguish three levels that should not be conflated: stabilising conditions that strengthen or support Presence; empirically resilient articulation-forms through which Presence is lived as here and now; and a transcendental limit-condition concerning the first-personal givenness of experience. Subtraction provides evidence for the resilience of here and now, but empirical resilience does not by itself establish constitutivity or transcendental necessity. I argue, on phenomenological rather than experimental grounds, that for-me-ness names the limit-condition within which any such spatial or temporal articulation can be experienced at all. The paper bridges operational Presence research with Husserlian givenness, Leibhaftigkeit, and image consciousness; rereads Bodyless Presence as exposing the resilience of here; rereads Bodyless Presentness as exposing the resilience of now; and develops a stratified account of Presence through Zahavi's pre-reflective self-awareness while taking Derrida's critique of self-presence seriously. It concludes by proposing immersive media as dissociation apparatuses for loosening conditions ordinarily coupled within experiential there-ness.

### I. INTRODUCTION: Present After Presence

Presence has been one of the central concepts in virtual reality, telepresence, immersive media, and XR research. It names, in its most familiar form, the user's sense of "being there" in a mediated environment (Sheridan, 1992; Steuer, 1992; Slater & Wilbur, 1997; Slater, 2009). Throughout this paper, "Presence" with a capital P refers to the operational construct developed within immersive-media research, whereas lowercase "presence" is reserved for broader phenomenological or ordinary senses of the term. The history of Presence research has often been organised around a broadly additive question: which factors enhance Presence? System-side immersion, wide field of view, stereoscopy, tracking fidelity, sensorimotor contingency, bodily representation, agency, ownership, interaction, realism, social responsiveness, and co-presence have all been examined as variables that may strengthen the experience of being situated within a mediated environment (Schuemie *et al*., 2001; Cummings & Bailenson, 2016; Skarbez, Brooks, & Whitton, 2017).

This additive orientation has been productive. It has clarified that Presence is not reducible to display quality alone; that system-side immersion and user-side Presence must be distinguished; that Place Illusion and other content-related dimensions can be experimentally manipulated; and that the user's body, action possibilities, and social context matter for the structure of mediated experience (Slater, 2009; Slater & Sanchez-Vives, 2016; Skarbez *et al*., 2017). Yet the same orientation also risks producing a tacit metaphysics of accumulation. If more embodiment, more agency, more interactivity, more realism, and more co-presence tend to increase Presence, it becomes tempting to treat these factors as constitutive of Presence itself.

The present paper proposes the opposite strategy. Rather than asking what must be added to enhance Presence, it asks what becomes disclosed when apparently indispensable conditions are progressively weakened or subtracted. This approach is not merely negative. Subtraction does not seek to show that embodiment, agency, co-presence, or reciprocity are unimportant. It asks a more precise question: which conditions stabilise or amplify Presence, and which conditions are minimally implicated in its possibility?

Two preceding lines of work motivate this question. The first is *Bodyless Presence*, which addressed immersive video as a boundary case between interactive virtual reality and conventional video. In immersive video, users may experience strong spatial and bodily Presence despite restricted locomotion, limited action, and the absence of a stable visible body. The phenomenon was described as a self-location-dominant state, in which self-location remains comparatively foregrounded while agency, ownership, and body-schema availability are attenuated (Toida, 2026a). The second is the framework of *Bodyless Presentness*, introduced in *Being and Time in XR: Other-Presentness Beyond Co-Presence* (Toida, 2026b), which addressed mediated others. It asked whether another individual must be physically and temporally co-present in order to be experienced as existing "here and now". Recorded performers, remote collaborators, spatial personas, and substitutional-reality events all suggest that bodily co-presence, real-time simultaneity, and reciprocal interaction can be weakened while other-presentness remains partially preserved (Suzuki, Wakisaka, & Fujii, 2012; Irlitti *et al*., 2023; Toida, 2026b).

The present paper does not simply combine these two frameworks. To combine them would be to enlarge the taxonomy: self-location here, other-presentness there, then perhaps event-presentness or world-presentness elsewhere. Such a move would remain within the additive logic it seeks to question. Instead, this paper proposes a more radical rereading. Bodyless Presence exposed the resilience of here: the spatial articulation of Presence as self-location within a world-horizon. Bodyless Presentness exposed the resilience of now: the temporal articulation through which recorded, asynchronous, or mediated events may still be experienced as presently unfolding. What these studies point towards is not a larger inventory of Presence components but a residual structure of givenness.

The central claim is as follows: Presence is a first-personal mode of givenness whose spatial and temporal articulations appear as here and now. Three different kinds of claim must, however, be distinguished. First, embodiment, agency, co-presence, reciprocity, simultaneity, and related factors can function as stabilising conditions: their presence may strengthen or support Presence, while their attenuation need not abolish it. Second, here and now are treated as articulation-forms: by articulation I mean the manner in which first-personal experience becomes spatially situated and temporally current, rather than an additional content or causal component of experience. Their persistence under subtraction establishes empirical resilience, not necessity. Third, for-me-ness concerns a different level of analysis. It names the pre-reflective first-personal character by which experience is given to someone at all (Zahavi, 1999, 2005, 2014).

These categories are not competing logical modalities. In particular, a transcendental condition need not be opposed to a necessary condition: a transcendental claim concerns a condition of possibility and may therefore also be a claim of necessity. Nor does "resilience" name a weaker kind of necessity. It names an empirical finding: a structure remains recognisable when selected supports are attenuated. The present paper therefore does not infer transcendental necessity from empirical survival. Subtraction establishes the resilience of here and now; phenomenological reflection motivates the further claim that first-personal givenness is the condition within which any such spatial or temporal articulation can appear.

On this basis, I propose for-me-ness as the transcendental limit-condition of Present After Presence. This is not a further residue isolated experimentally after here and now. Every act of subtraction considered in this paper already occurs within first-personal experience. For-me-ness therefore marks the limit at which empirical subtraction gives way to transcendental reflection: to subtract it would no longer produce a modified experience whose structure could be examined, but would remove the very field in which anything could appear as here or now.

This claim does not restore a metaphysics of pure self-presence. The argument accepts the post-Derridean lesson that presence is temporally differentiated, mediated, and never free from trace, while maintaining that such mediation does not eliminate the minimal first-personal character of experience.

The argument proceeds in six steps. Section 2 clarifies the bridge between operational Presence research and phenomenological analyses of givenness, especially Husserl's concepts of *Leibhaftigkeit*, *Gegenwärtigung*, *Vergegenwärtigung*, and image consciousness. Section 3 rereads Bodyless Presence as the exposure of here. Section 4 rereads Bodyless Presentness as the exposure of now. Section 5 examines the limit of subtraction through the relation among for-me-ness, here, and now, engaging Zahavi and Derrida. Section 6 develops the methodological implications of this account by treating immersive media as dissociation apparatuses: they allow the stabilising conditions of here and now to be varied without treating for-me-ness itself as an experimental variable. Section 7 concludes by defining Present After Presence as the question of what becomes disclosed when the conditions thought to produce Presence have been progressively weakened or removed.

## II. THEORETICAL BACKGROUND: Operational Presence, Leibhaftigkeit, and Image Consciousness

Before the preceding subtraction studies can bear upon a phenomenological question, two vocabularies must be kept distinct. In XR research, Presence names a psychological experience of being situated within a mediated environment (Slater & Wilbur, 1997; Slater, 2009; Skarbez *et al*., 2017). Husserlian phenomenology does not employ "presence" in this operational sense. It distinguishes modes in which objects and events are given: perceived, imagined, remembered, represented, or otherwise intended. The argument below does not identify these vocabularies or treat one as the operationalisation of the other.

One relevant Husserlian concept is Leibhaftigkeit, the character of perceptual givenness in which a spatial object is given "in the flesh" rather than merely represented in its absence (Husserl, 1982, 2001). The object is given through perspectival profiles and remains inexhaustible by any single appearance. This concept concerns the mode of givenness of the perceived object; it should not be confused either with XR Presence as a psychometric construct or with the temporal sense of an event's being experienced as occurring now.

The bridge proposed here is therefore methodological rather than identificatory. Mediated environments create cases in which representational status, spatial situatedness, bodily possibility, and temporal currentness can cease to vary together. XR Presence measures provide empirical access to some of these disturbances, while phenomenological concepts provide a vocabulary for describing what kind of experiential relation has changed. The purpose is not to reduce Leibhaftigkeit to Presence, but to use technologically produced boundary cases to ask how mediated environments shift between being encountered as represented objects and being lived as fields within which experience is situated.

This contrast approaches the distinction Husserl marks between Gegenwärtigung and Vergegenwärtigung: between presentive givenness and modes in which something absent is re-presented. The correspondence is not exact. A technological medium cannot be assigned to one intentional mode solely by virtue of being live, recorded, or simulated. A live video may remain experienced as a represented object; a recorded immersive environment may nevertheless function as the field in which the user is currently situated. The point is not that media format determines intentional mode. The point is that technological media can reorganise the relation between representation and experiential currentness.

Mediated experience also introduces a complication that cannot be captured by the distinction between presentation and re-presentation alone. Immersive video and virtual environments are not encountered independently of their status as technologically produced images. Husserl's analyses of image consciousness distinguish the physical image carrier, the image-object that appears through it, and the image-subject that is depicted (Husserl, 2005). A recorded or simulated environment cannot therefore be classified simply as re-presentation because of its technical origin. The relevant question is how the image is lived within experience.

In paradigmatic picture viewing, the image remains apprehended as an image: the viewer encounters a represented subject through an image-object that remains distinct from the surrounding perceptual world. In immersive XR, by contrast, the physical carrier and image-object may recede from thematic awareness, while the depicted environment begins to function as the field within which perception, orientation, and possible action unfold. This does not establish that the image-subject has become leibhaftig in Husserl's full sense. It indicates that the mediated environment is no longer lived primarily as an object of representation, but as an experientially inhabited field.

This restricted bridge is sufficient for the argument that follows. The subtraction studies do not show that the lived body is irrelevant to perceptual experience, nor that technological mediation reproduces originary perception without remainder. They show that the experiential force of being situated there may remain comparatively resilient even when several conditions ordinarily associated with embodiment, agency, or physical co-presence are attenuated. Immersive media are philosophically significant not because they duplicate Leibhaftigkeit, but because they permit conditions ordinarily fused within experiential there-ness to be partially decoupled.

The analysis developed in this paper redescribes this resilience through distinctions that are not presented as Husserl's own. The first-personal character of experience will be described as its for-me dimension, while here and now will name the spatial and temporal articulations through which such givenness becomes experientially determinate. Immersive media do not reveal the entirety of Leibhaftigkeit. They provide a technical setting in which some of the conditions ordinarily supporting experiential there-ness can be weakened, thereby making their relations available for renewed phenomenological analysis.

## III. FIRST SUBTRACTION: The Resilience of Here

The first subtraction concerns embodiment. A long tradition of VR and embodiment research has shown that bodily representation, sensorimotor contingency, agency, body ownership, and possible action profoundly shape mediated experience (Slater & Wilbur, 1997; Slater, 2009; Kilteni, Groten, & Slater, 2012; Blanke & Metzinger, 2009). The virtual body is not merely an object added to the scene. It can organise proprioceptive expectation, peripersonal space, motor possibility, and the sense that the user is located within the virtual world. Conversely, disruptions to tracking, latency, or bodily congruence can weaken the sense of being there.

This body-centred account is compelling. Yet immersive video introduces a boundary case. In 180° or 360° stereoscopic video, the user may experience spatial Presence while lacking many of the conditions associated with interactive VR. Locomotion is restricted. Environmental interaction is minimal or absent. The user typically does not see a stable virtual body corresponding to their own. Agency over the environment is limited, and body ownership is weak or unavailable. Nevertheless, the user may still feel situated at the recorded viewpoint. They may feel that events occur around them rather than merely before them. They may flinch when the camera is approached, feel exposed when directly addressed, or experience movement and impact around the camera as events concerning their own position.

Bodyless Presence was proposed to describe this boundary condition (Toida, 2026a). The term does not mean that the user has no body. Nor does it imply that the body schema disappears. Rather, it describes a configuration in which the availability of body schema, agency, and ownership is attenuated while self-location becomes comparatively dominant. The crucial point is not that embodiment is dispensable. It is that self-location can persist under conditions in which several bodily stabilisers of Presence are weakened.

The present paper rereads this result as the exposure of here. In ordinary embodied action, here is densely supported. The lived body provides an orientation zero, a field of possible movement, a distinction between reachable and unreachable space, and a structure of practical relevance (Merleau-Ponty, 2012; Gallagher, 2005). Because these structures normally co-occur, it is easy to treat here as inseparable from bodily agency or ownership. Immersive video loosens this density. It can preserve a viewpoint, orientation, stereoscopic depth, and head-contingent looking while weakening action, ownership, and bodily visibility. The result is not pure disembodied existence, but a partial decoupling through which the spatial articulation of Presence becomes visible.

Here, in this sense, is not exhausted by a geometric coordinate. A camera position can be specified from a third-person perspective without being lived as anyone's position. A lived here, by contrast, functions as an orientational centre: things appear in front of or behind me, near or far from me, reachable or unreachable from where I am situated. This is what is meant here by calling here "place-like" rather than merely locational, in the phenomenological sense that place is lived as an oriented field rather than specified merely as an abstract position (Casey, 1993; Trigg, 2012). The claim is not that such organisation is disembodied. On the contrary, it remains bodily implicated through orientation, posture, peripersonal organisation, and possible movement even when a visible avatar or free positional locomotion is unavailable. Bodyless Presence therefore names not the absence of the lived body, but a technologically unusual dissociation between bodily situatedness and some of its ordinary visual and action-related supports. A mediated field becomes a lived here when the captured viewpoint functions not merely as a coordinate from which an image was recorded, but as the centre around which the experienced field is organised.

Empirical work on immersive video and filming distance supports this interpretation. In a study manipulating filming distance during a 180° stereoscopic live-action performance, high-Presence and low-Presence conditions differed not only in visual composition but in Spatial Presence, Self-Location, Possible Actions, Social Presence, and Observability (Toida *et al*., 2026). The near-distance condition did not merely make performers appear larger. It positioned the viewer within a different pseudo-interpersonal configuration. In this sense, filming distance functioned as a design variable for here: it specified where the viewer was experientially located in relation to the performers.

The philosophical significance of this result is limited but important. It does not prove that here is a necessary or transcendental condition of Presence. Rather, it shows that here is empirically resilient: its spatial articulation can remain recognisable when several bodily stabilisers are attenuated. Nor does this imply that body, agency, or ownership are irrelevant. The first subtraction therefore yields a negative and a positive lesson. Negatively, Presence should not be reduced to bodily ownership, agency, or interactive possibility. Positively, when these are weakened, the spatial articulation of Presence as here may remain.

This reinterpretation also requires self-critique. The earlier notion of a self-location-dominant state remains useful as a local description of immersive video. However, it should not be elevated into a general theory of Presence. Self-location is not the final ground of Presence. It is the way first-personal givenness becomes spatially determinate under certain conditions. To absolutise self-location would merely replace one additive model with another. The present paper therefore treats Bodyless Presence not as the discovery of a new ingredient, but as the first subtraction through which the resilience of here became visible.

## IV. SECOND SUBTRACTION: The Resilience of Now

The second subtraction concerns time. In everyday interpersonal experience, physical co-presence, spatial proximity, bodily orientation, gaze, responsiveness, and temporal simultaneity normally co-occur and mutually support one another. Another person is normally encountered as present because they are there with me, now, available for mutual orientation and potential response. This density has shaped research on social presence, co-presence, telepresence, joint attention, and interaction (Biocca, Harms, & Burgoon, 2003; Frischen, Bayliss, & Tipper, 2007; Irlitti *et al*., 2023). It has also shaped common intuitions about mediated others: recorded people belong to the past; live remote people belong to the present; fictional or parasocial figures belong to representation.

XR complicates this intuition. Spatial Persona and volumetric telepresence can weaken physical co-location while preserving forms of shared spatiality. Substitutional Reality can present recorded scenes in ways that are experienced as currently unfolding (Suzuki *et al*., 2012). Immersive video can present recorded performers as if they were near the viewer's bodily position. Parasocial interaction can generate relational responses across long temporal distances (Horton & Wohl, 1956; Tukachinsky & Stever, 2019). The question is not simply whether mediated others can evoke social responses. That has long been established. The more precise question is whether and how mediated events acquire currentness.

Bodyless Presentness was formulated to address this problem (Toida, 2026b). It described conditions under which another individual continues to be experienced as existing "here and now" despite attenuated bodily co-presence and weakened real-time simultaneity. The original framework identified candidate structures such as action predictability, attentional legibility, spatial configuration, expectation of responsiveness, and causal coherence. These structures remain useful. They describe ways in which mediated others may become socially and perceptually legible even when bodily co-presence or reciprocity is weakened.

However, the present paper rereads the philosophical yield of Bodyless Presentness differently. The primary residual is not other-presentness itself. Other-presentness is a complex phenomenon stabilised by social, perceptual, spatial, causal, and affective structures. Its conditions can be analysed, manipulated, and compared, but it is not the minimal residual of Presence. What the second subtraction exposes more directly is now. Recorded, delayed, staged, or asynchronous events may be known as non-live and yet be experienced with a form of currentness. They may not be believed to be happening in objective real time, but they can nevertheless appear as events unfolding in the experiential present.

This distinction matters. Objective simultaneity and experiential currentness are not identical. A live video call may be experienced as alienated, flat, or observational; a recorded immersive scene may be experienced as surrounding and unfolding. Conversely, knowledge that something is recorded may weaken claims of actual reciprocity without entirely eliminating the temporal force of its appearance. The now relevant to Presence is therefore not clock time. It is not mere liveness. It is the temporal articulation through which events are experienced as belonging to the ongoing stream of experience.

Husserl's analyses of internal time-consciousness are indispensable here. The living present is not a punctual now-point. It includes retention and protention: the just-past retained within the present and the immediate future anticipated within it (Husserl, 1991). A melody, movement, or gesture is not present as an isolated instant but as a temporally thick unfolding. This thickness allows recorded media to present events that, although objectively past, are experientially encountered as unfolding now. The question is not whether the event is live. The question is whether the event is integrated into the temporal flow of experience as current.

The claim here is deliberately restricted to the currentness of the mediated event. It does not treat the event's temporal givenness as identical with the longitudinal self-givenness of consciousness itself. In Husserlian terms, the argument concerns how an intended event is encountered within an ongoing temporal field, not whether a "now" can somehow be subtracted from the first-personal temporal flow that makes such encounter possible. Accordingly, the second subtraction does not yield a now independent of for-me-ness. It shows something narrower but still important: objective liveness is not necessary for an event to be lived as currently unfolding within that first-personal flow.

Substitutional Reality provides one extreme case. By blending present perception with recorded scenes, it can allow past events to be experienced as if they were currently occurring (Suzuki *et al*., 2012). Immersive video provides a less radical but more widely available case. The viewer may know that the performance was recorded, yet the combination of stereoscopy, spatial audio, head-contingent looking, viewpoint proximity, and event continuity can allow the recorded performance to function as a present field of experience. This is not deception in the simple sense. It is a dissociation between technical recordedness and experiential currentness.

Being and Time in XR originally framed this dissociation in terms of other-presentness. This was appropriate for its immediate aim: to theorise how mediated others can be experienced as present beyond co-presence. Yet when reread as part of the present argument, its deeper contribution lies in exposing the resilience of now. Bodyless Presentness weakened real-time simultaneity, reciprocity, and bodily co-presence; what remained was not merely the social meaning of the other, but the temporal articulation through which the represented event continued to appear as currently unfolding.

As with here, this resilience should not be confused with constitutivity. The persistence of now under temporal subtraction does not prove that now is an independent necessary ingredient of Presence. It shows that the temporal articulation of first-personal givenness can survive the attenuation of several temporal stabilisers. Liveness, reciprocity, latency reduction, and synchronous interaction remain powerful stabilising conditions. They can intensify, clarify, and socially validate currentness. But they are not identical with currentness itself.

The evidential status of this second subtraction is more programmatic than that of the first. Whereas the spatial subtraction can be supported by direct empirical manipulation of filming distance and self-location-related Presence measures, the temporal subtraction presently relies more heavily on conceptual analysis and prior work on recorded, asynchronous, and substitutional-reality experience. The symmetry between the first and second subtraction is therefore theoretical rather than evidentially complete. Future work should test the resilience of now through more direct manipulations of liveness, delay, replay, and temporal framing.

This reinterpretation also reframes the five structures proposed in Bodyless Presentness. Action predictability, attentional legibility, spatial configuration, expectation of responsiveness, and causal coherence should no longer be described as constitutive structures of other-presentness in a strong sense. They are better understood as stabilising structures. They help mediated others and events acquire social legibility and temporal currentness. They do not constitute Presence at its most minimal level. The second subtraction therefore yields the temporal counterpart to the first: Bodyless Presence exposed the resilience of here; Bodyless Presentness exposed the resilience of now.

## V. THE LIMIT OF SUBTRACTION: For-Me-Ness, Here, and Now

The preceding sections expose a methodological reversal. Empirically, subtraction first makes here and now visible because immersive media can selectively disturb the conditions that ordinarily support spatial situatedness and temporal currentness. Phenomenologically, however, neither can be understood independently of the first-personal field within which anything appears as here or now. The order of empirical disclosure is therefore here and now before for-me-ness; the proposed order of phenomenological foundation runs in the opposite direction.

These should not be treated as two sealed realms, one empirical and the other transcendental. Empirical variation can disclose dependencies that motivate transcendental reflection, just as phenomenological variation can clarify what is essential by considering what can and cannot vary. The distinction concerns what follows from each procedure. Empirical subtraction can show that a support is dispensable in a particular configuration, or that an articulation remains resilient under attenuation. It cannot by itself establish that whatever survives is transcendentally necessary.

For-me-ness occupies a different position. It is not another residue isolated after body, agency, liveness, or reciprocity have been progressively removed. Every subtraction considered here already occurs as a modification of someone's experience. I therefore propose for-me-ness as the transcendental limit-condition of the procedure: not because an experiment has shown it to survive every possible subtraction, but because removing first-personal givenness would remove the field within which subtraction could appear as an experiential variation at all.

"Transcendental" is used here in the phenomenological sense of a condition of possibility. It is therefore not opposed to necessity. Nor is it an alternative to empirical description. The claim is instead stratified: stabilising conditions are empirically variable supports; here and now are articulation-forms whose resilience can be examined empirically; for-me-ness is proposed phenomenologically as the condition within which any such articulation is given. By articulation, I mean the way first-personal givenness becomes determinate as spatially situated and temporally current, not an independent substance, component, or causal mechanism.

This formulation depends on a precise account of for-me-ness. For-me-ness does not mean reflective self-consciousness. It does not mean that I perceive myself as an object. It does not mean that I judge the experience to be mine. Nor is it equivalent to bodily ownership. Rather, it refers to the pre-reflective character of experience by which experience is lived from a first-personal perspective (Zahavi, 1999, 2005, 2014; Gallagher & Zahavi, 2008). This formulation also stands in proximity to Sartre's account of pre-reflective consciousness, while the present argument follows Zahavi in treating for-me-ness as a minimal structure of experiential givenness rather than as a theory of egoic selfhood (Sartre, 1956, 2004; Zahavi, 1999, 2005). To experience pain, colour, movement, space, music, or another person is not first to encounter a neutral content and then infer that it belongs to me. The experience is already given as lived. This does not make the self transparent or substantial; it marks the minimal subjectivity of experiencing.

This commitment is not intended as an uncontroversial grammatical truth. It aligns the present argument with phenomenological accounts of pre-reflective mineness rather than with no-self or self-model theories that treat the self as a constructed representational model, or with higher-order accounts that locate consciousness in reflective or meta-representational structures (Metzinger, 2003; Rosenthal, 2005; Zahavi, 1999, 2005). The claim is therefore substantive but deliberately minimal: the argument does not require a substantial self, a transparent ego, or an objectified inner subject, but only the non-anonymous givenness of experience.

A self-model theorist might object that this merely names the explanandum rather than explaining it. On such a view, mineness, self-location, ownership, and agency are not primitive structures of experience but representational effects generated by an integrated self-model; higher-order accounts may likewise argue that consciousness depends on a further representational relation by which a mental state becomes available as conscious. The present argument does not deny the explanatory force of such accounts at the level of cognitive architecture. It grants that bodily ownership, agency, autobiographical selfhood, and even many forms of self-recognition may be constructed, modelled, or reflectively mediated.

The disagreement concerns the level at which the claim is made. For-me-ness is not introduced as an additional represented content, inner owner, or subpersonal mechanism competing with a self-model. It names the dative structure in virtue of which any content, including a self-model, is experientially manifest at all. A self-model may explain how the subject is represented within experience; it does not by itself show that experience can be phenomenally manifest without being given from any perspective. Likewise, a higher-order representation may explain reflective availability or reportability, but it presupposes that the relevant state is lived or manifest in some manner. The point is therefore not that self-model theory is false, but that the constructedness of self-content does not entail the anonymity of experience. What can be subtracted are particular self-representations, bodily models, ownership structures, and reflective identifications; what cannot be subtracted without eliminating experience is the non-anonymous givenness within which such structures appear.

Derrida's critique of presence cannot be avoided at this point. Any claim about presence, the living present, or self-givenness risks appearing to restore the very structure that Derrida deconstructed in his reading of Husserl: the privilege of self-presence, the purity of the now, and the supposed immediacy of voice to itself (Derrida, 1973). The present argument does not seek to bypass this critique by saying that its for-me-ness is merely "minimal" and therefore immune.

That would be insufficient. Derrida's critique reaches precisely into the living present, showing how retention, trace, and temporal difference inhabit what might otherwise be taken as pure self-presence.

The response is not to deny this critique, nor to protect for-me-ness by placing it beneath the reach of *différance*. If for-me-ness meant an immediate coincidence of consciousness with itself, Derrida's critique would be decisive. But this is not the sense intended here. For-me-ness does not name a pure self-presence that would precede temporal differentiation. It names the non-objectifying fact that experience is not first given as anonymous content to which a subject is subsequently added. Even when experience is internally marked by retention, protention, trace, and delay, it is still lived from somewhere, or better, as lived at all. The first-personal character of experience is therefore not an island of purity immune to mediation, but the minimal dative structure within which mediation, absence, and temporal differentiation can appear.

Immersive media make this point methodologically salient. Immersive video, recorded performance, latency, replay, and substitutional reality all present experiences that are technically mediated, temporally displaced, and structured by trace rather than pure immediacy. Yet such experiences are not thereby given as anonymous contents. They are still lived from a first-personal orientation, even when their here and now are technologically reconfigured. In this respect, immersive media do not refute Derrida by restoring pure presence. They provide boundary cases in which mediated, delayed, and trace-structured experience nevertheless remains first-personally given.

This is why Zahavi's account remains useful without requiring a return to metaphysical self-presence (Zahavi, 1999, 2005). Pre-reflective self-awareness is not an inner perception of a self-object, nor a transparent possession of oneself. It is the intrinsic mineness or for-me character of experience prior to reflective identification. Derrida's critique prevents this structure from being misunderstood as self-identity without difference; it does not require the stronger conclusion that experience is first-personally nowhere. The present argument therefore accepts the deconstruction of pure presence while retaining a thinner claim: experience need not coincide with itself in order to be given first-personally.

Now must likewise be defined carefully. The now relevant to Presence is not a punctual instant purified of difference. It is a thick present structured by retention and protention. It includes temporal spread, anticipation, decay, and horizon. The event that appears now is already internally differentiated. Thus, the second subtraction does not recover a pure present behind mediation. It shows that mediated events can acquire currentness within the temporal flow of experience, even when objective simultaneity is weakened.

Here must also be defined non-naively. The here relevant to Presence is not a physical coordinate. It is the spatial articulation of first-personal givenness within a world-horizon. It includes orientation, proximity, possible movement, bodily implication, and the difference between centre and periphery. Merleau-Ponty's account of the body as our general medium for having a world remains essential here (Merleau-Ponty, 2012). Yet Bodyless Presence shows that several bodily stabilisers of here can be weakened while the articulation remains. Here is therefore not reducible to visible body ownership or overt motor agency. It is the spatial form in which for-me-given experience becomes oriented.

This yields the central distinction of the paper. Embodiment, agency, co-presence, reciprocity, and simultaneity function as stabilising conditions whose contribution can vary empirically. Here and now exhibit empirical resilience: they remain articulated across selected subtractions, although that resilience does not establish them as necessary conditions. For-me-ness occupies a different logical position. It is proposed as a transcendental limit-condition because every spatial or temporal articulation examined here is already given within a first-personal field. This is a phenomenological claim about the condition of possibility of experience, not an empirical inference from whatever happens to survive a sequence of manipulations.

The error to avoid is therefore twofold. Additive reasoning can mistake what strengthens Presence for what constitutes it. Subtractive reasoning can make the inverse mistake, treating whatever survives attenuation as necessary merely because it survives. Present After Presence rejects both moves. Stabilisation, resilience, and transcendental conditionality answer different questions: what supports an experience, what persists when selected supports are weakened, and what must be presupposed for the experience to be given at all.

This structure also clarifies the status of the earlier concepts. Self-location-dominant state and the five structures of other-presentness are not abandoned, but reclassified. They are local models describing how Presence becomes spatially or socially stabilised under particular media conditions. They do not name the residual limit-condition of Presence. Bodyless Presence and Bodyless Presentness therefore function retrospectively as subtraction studies: not because they discovered new constitutive ingredients, but because they exposed the forms through which first-personal givenness remains spatially and temporally articulated after stabilising conditions are weakened.

Present After Presence can therefore be defined as the question of what remains present once the additive model of Presence has been put into question. It does not name a return to pure immediacy or a final empirical residue exposed after mediation has been stripped away. It names a double movement: empirical subtraction makes the resilience of here and now visible, while phenomenological reflection asks what must already be presupposed for anything to be lived as here or now at all. The answer proposed here is first-personal givenness, or for-me-ness.

## VI. METHODOLOGICAL IMPLICATIONS: Immersive Media as Dissociation Apparatuses

The preceding argument is philosophical, but its methodological force depends on a distinctive feature of immersive media: they can function as dissociation apparatuses. They do not provide evidence for for-me-ness in the way they provide evidence for the resilience or collapse of here and now. Rather, they make it possible to loosen conditions that ordinarily co-occur within everyday experience: embodiment, agency, ownership, co-presence, reciprocity, simultaneity, and spatial orientation. In this respect, immersive media are valuable not because they produce Presence more strongly, but because they allow the stabilising conditions of being-there to be weakened, displaced, or recombined.

This diagnostic approach also continues a line of work in which temporal perturbations of self-generated auditory feedback were used to examine the behavioural and neural limits of sensorimotor integration. Earlier studies showed that subjective simultaneity between action and auditory feedback can be recalibrated within a limited delay range, and that delayed feedback elicits ERP components associated with conscious delay detection and reduced agency (Toida, Ueno, & Shimada, 2014, 2016). This earlier work does not provide evidence for the present account as a whole. It provides a methodological precedent: temporal perturbation can be used to examine the limits within which experience preserves or loses a sense of sensorimotor coherence.

Here and now, by contrast, can be investigated through their stabilising conditions. The relevant method is not clean removal but attenuation and resilience. In embodied and spatial Presence, here may be stabilised by visible body representation, viewpoint-body congruence, sensorimotor contingency, field of view, stereoscopy, viewpoint height, locomotion, possible action, and peripersonal spatial cues. In temporal presentness, now may be stabilised by liveness, continuity, response latency, temporal framing, causal coherence, event pacing, conversational turn-taking, and knowledge of whether an event is live or recorded. Taken together, these manipulations define a programme of patterned attenuation rather than component isolation, extending the experimental logic of Presence research towards phenomenological variation (Slater, 2009; Skarbez *et al.*, 2017; Varela, 1996).

An experimental phenomenology of here would vary these stabilisers while measuring the persistence, weakening, or transformation of self-location. For example, immersive video can manipulate filming distance, viewpoint height, camera motion, direct address, visual access to one's body, or the continuity between head movement and visual update (Toida *et al*., 2026). Interactive VR can manipulate body visibility, agency, locomotion, and sensorimotor congruence. The question is not simply whether Presence scores increase. It is how the field becomes organised as a here: whether the user feels located within the environment, whether events are experienced as occurring around the user, whether disruptions foreground the physical body, and whether recovery re-stabilises self-location.

An experimental phenomenology of now would vary temporal stabilisers while measuring experiential currentness. Live and recorded immersive video can be compared while holding visual content constant. Response latency can be manipulated in remote collaboration. Recorded events can be temporally framed as live, archival, replayed, fictional, or delayed. Event continuity can be disrupted or preserved. Substitutional Reality can introduce recorded scenes into present perception. The central question is not whether participants believe the event is objectively live, but whether it functions as currently unfolding within experience.

This distinction between belief and currentness is crucial. A user may know that an immersive video is recorded and nevertheless experience it as unfolding now. Conversely, a live remote event may be known as live while failing to acquire experiential currentness because of latency, poor spatial configuration, weak social orientation, or disrupted continuity. Future studies should therefore distinguish explicit temporal belief from experienced currentness. Measures may include self-report, behavioural orientation, anticipatory responses, gaze dynamics, physiological responses, and qualitative phenomenological interviews.

The programme also requires methodological caution. Here and now should not be treated as cleanly orthogonal variables. Spatial and temporal articulation are intertwined. A disruption in tracking may affect both the sense of here and the flow of now. A latency manipulation may alter temporal currentness while also weakening co-presence and action possibility. Similarly, gaze, direct address, and interpersonal distance may affect spatial self-location, social presence, and temporal engagement at once. The appropriate model is not independent component isolation but patterned attenuation.

This is what gives immersive media their methodological value. In everyday experience, embodiment, agency, co-presence, simultaneity, and world-horizon are normally interdependent and jointly organised. Such media do not make these dimensions fully separable, but they can loosen their coupling. They allow the researcher to ask which stabilisers can be weakened without eliminating here, which temporal disruptions undermine now, and which combinations produce breakdowns in Presence. Such breakdowns are not methodological failures. They are diagnostic. They reveal the conditions under which articulation collapses, shifts, or re-stabilises.

The resulting programme differs from standard presence engineering. Presence engineering asks how to maximise Presence. Experimental phenomenology asks how Presence is articulated, destabilised, and reconstituted. It treats immersive media not merely as devices for producing experience, but as technical situations in which the structure of experience can be varied. This is why immersive media remain methodologically relevant to the present argument even though its central claim is phenomenological. They are not objects of philosophical celebration; they are the technical conditions that make subtraction experimentally available. The programme also differs from a simple return to Husserl. Phenomenology traditionally proceeds by description, reduction, and eidetic variation. Immersive media cannot replace these methods, nor do they turn phenomenological description into a psychometric variable. Rather, in a spirit closer to experimental and neurophenomenological programmes, they provide a technical analogue to variation: a way of weakening, separating, and recombining conditions that are normally fused within everyday experience (Varela, 1996). In this sense, immersive media can support a form of experimental phenomenology. They do not empirically prove transcendental claims. They supply controlled boundary cases through which such claims can be sharpened, constrained, and revised.

## VII. CONCLUSION: After the Additive Model

The argument of this paper can be summarised as a series of reclassifications. Embodiment, agency, ownership, co-presence, reciprocity, and temporal simultaneity are not dismissed. They are reclassified as stabilising and amplifying conditions. They make Presence stronger, clearer, more socially validated, more action-oriented, and more durable. But their importance does not entail that they are constitutive in the strongest sense.

Bodyless Presence weakened bodily stabilisers and exposed the resilience of here. Bodyless Presentness weakened co-presence, reciprocity, and real-time simultaneity and exposed the resilience of now. Present After Presence places these two subtractions within a broader phenomenological structure. Here and now describe the spatial and temporal articulation of Presence; for-me-ness names the first-personal givenness within which either articulation can be lived at all.

The argument does not infer this hierarchy directly from empirical survival. Stabilising conditions, resilient articulation-forms, and transcendental conditionality answer different questions. The first concerns what supports or intensifies an experience. The second concerns what remains recognisable when selected supports are weakened. The third concerns what must be presupposed for such an experience to be possible at all. On phenomenological grounds, I have argued that for-me-ness occupies this third position. It is not the final object left behind by subtraction, but the limit at which subtraction ceases to yield another modified experience and instead confronts the first-personal givenness presupposed by every modification.

The title Present After Presence therefore has a double sense. It refers first to what becomes present after the additive research programme has been interrupted: after the assumption that Presence is best understood by accumulating its contributing conditions. It also refers to the experiential present after the critique of pure self-presence, pure immediacy, and the punctual now. What remains is not metaphysical presence restored, but mediated, horizonal, temporally thick, first-personal givenness.

This is why Derrida does not stand outside the argument. The paper does not attempt to defend a pure present against Derrida's critique. It accepts that the living present is internally differentiated, that presence is never free from absence, and that self-presence cannot be understood as transparent possession. Yet it also maintains, with Zahavi, that the critique of robust self-presence need not eliminate minimal for-me-ness. Experience need not be self-transparent in order to be first-personal.

The paper therefore repositions immersive media. They are not merely technologies for producing stronger Presence, nor media for reproducing reality. Their philosophical significance lies in the possibility of subtraction: they loosen the dense co-occurrence of body, space, time, action, and otherness, thereby making visible relations that everyday experience normally conceals.

Present After Presence is therefore not a theory of more Presence. It asks what becomes disclosed when Presence is reduced, strained, and partially deprived of the conditions thought to sustain it. Subtraction exposes the resilience of here and now without turning that resilience into necessity. At the point where subtraction can no longer be conceived as a variation of experience without presupposing that the variation is given to someone, the argument passes from empirical manipulation to transcendental reflection. Present After Presence names this double movement: the experimental weakening of the supports of being-there and the phenomenological attempt to articulate what it means for anything to remain present for someone at all.

## ACKNOWLEDGMENT

Much of what made this text possible remains outside the text. I am grateful to those who were present in its making, whether through discussion, criticism, encouragement, or silence.